**Predictability as a probe of manifest and latent physics: the case of atomic scale structural, chemical, and polarization behaviors in multiferroic Sm-doped BiFeO$_3$**

Maxim Ziatdinov,[1,6,*] Nicole Creange,[2] Xiaohang Zhang,[3] Anna Morozovska,[4] Eugene Eliseev,[5] Rama K. Vasudevan,[1] Ichiro Takeuchi,[3] Chris Nelson,[1] and Sergei V. Kalinin[1,†]

[1] The Center for Nanophase Materials Sciences, Oak Ridge National Laboratory, Oak Ridge, TN 37831

[2] Department of Materials Science and Engineering, North Carolina State University, Raleigh, NC 27606

[3] Department of Materials Science and Engineering, University of Maryland, College Park, MD 20742

[4] Institute of Physics, National Academy of Sciences of Ukraine, 46, pr. Nauky, 03028 Kyiv, Ukraine

[5] Institute for Problems of Materials Science, National Academy of Sciences of Ukraine, Krjijanovskogo 3, 03142 Kyiv, Ukraine

[6] Computational Sciences and Engineering Division, Oak Ridge National Laboratory, Oak Ridge, TN 37831

The predictability of a certain effect or phenomenon is often equated with the knowledge of relevant physical laws, typically understood as a functional or numerically derived relationship between the observations and known states of the system. Correspondingly, observations inconsistent with prior knowledge can be used to derive new knowledge on the nature of the system or indicate the presence of yet unknown mechanisms. Here we explore the applicability of Gaussian Processes (GP) to establish predictability and uncertainty of local behaviors from multimodal observations, providing an alternative to this classical paradigm. Using atomic resolution Scanning Transmission Electron Microscopy (STEM) of multiferroic Sm-doped

[*] ziatdinovma@ornl.gov
[†] sergei2@ornl.gov

BiFeO$_3$ across a broad composition range, we directly visualize the atomic structure and structural, physical, and chemical order parameter fields for the material. GP regression is used to establish the predictability of the local polarization field from different groups of parameters, including the adjacent polarization values and several combinations of physical and chemical descriptors, including lattice parameters, column intensities, etc. We observe that certain elements of microstructure including charged and uncharged domain walls and interfaces with the substrate are best predicted with specific combinations of descriptors, and this predictability and their associated uncertainties are consistent across the composition series. The associated generative physical mechanisms are discussed. It is also found that certain parameter combinations tend to predict the orthorhombic phase in the cases where rhombohedral phase is observed, suggesting potential role of clamping and confinement phenomena in phase equilibrium in Sm-BiFeO$_3$ system close to morphotropic phase boundary. We argue that predictability and uncertainty in observational data offers a new pathway to probe the physics of condensed matter systems from multimodal local observations.

Materials with competing order parameters represent one of the most fascinating objects in modern physics. Examples range from ferroelectric and ferroelastic materials in the vicinity of symmetry-incompatible morphotropic phase boundaries,[1, 2,73,74] ferromagnetic superconductors and materials such as $UGe_2$,[3-7] charge separated oxides,[8-10] and many others. The competition often gives rise to the complex non-uniform ground states[11-14] with many degenerate energy minima, and often yields exceptional functional responses to external stimuli, ranging from giant electromechanical coupling and dielectric constants in ferroelectric relaxors,[15-20] to large volume changes in phase change memory alloys,[21-23] and giant magnetoresistance in nanoscale phase-separated manganites.[13, 24-27] In many cases, these phenomena manifest in doped materials, where local atomic disorder can significantly affect competition between the order parameters, as exemplified e.g. by ferroelectric relaxors or cation-substituted manganites, even though in certain cases the phase coexistence can be induced by external stimulus such as magnetic or electric fields, or uniaxial or hydrostatic pressure or strain.[28,75,7629]

These materials offer a challenge in terms of developing models capable of describing and predicting their behavior. For homogeneous materials, the natural description on the mesoscale level can be explored within Landau theory, where the properties of interest are associated with the appropriate order parameter and the energy of the system is represented as its functional. For ferroelectrics, ferroelastics, and multiferroics the relevant description can be given by Ginzburg-Landau theory with a real order parameter,[30-32] whereas complex order parameter fields can be used for description of superconductors.[33-36] The mesoscopic order parameter based approach naturally allows for classification of phase transitions, identification of topological defects, etc.[37] However, the applicability of these mesoscopic models for atomically disordered systems remains a topic of continuous controversy and has been for several decades.[38-40]

Alternatively, the physics of these materials can be explored using lattice models, with the materials structure or functionality represented via a collection of local spins interacting via local interactions,[41, 42] potentially mediated by long-range depolarization fields. These models can be explored numerically or in special cases analytically to yield corresponding phase diagrams, susceptibilities, response functions, and other macroscopic functionalities.[41, 42] On a local level, these models can give rise both to ordered and disordered ground states, offering considerably deeper insight into the physics of the material.[12] Yet, the key issue both for the mesoscale and lattice type models is comparison with experiment, both as a way to explore the applicability of the model, extract relevant materials parameters, and enable forward prediction and thus materials optimization. For spatially uniform ground states, the combination of diffraction techniques with macroscopic measurements offers comprehensive insight into materials functionality, since the mesoscopic order parameter fields provide a sufficient description of the system. This however is not the case for disordered systems or systems with competing and spatially inhomogeneous ground states, since macroscopically averaged scattering data necessarily loses the relevant information.[43, 44]

Until recently, the analysis of such systems was possible only via complex diffraction techniques, where the analysis of the correlated disorder allowed reconstruction of possible microscopic models.[44-46] However, the associated inverse problems generally lead to multiple possible solutions, necessitating additional structural and compositional information. Recently, the emergence of high-resolution aberration corrected electron microscopy has provided an opportunity to visualize these systems at the atomic level and detect minute distortions of local atomic structures that can serve as a proxy for local order parameter fields. Namely, site variations of symmetry can be extracted from the atomic coordination neighborhood to identify changes in local symmetry class, i.e. different phases [47, 48], or quantify the strength of symmetry lowering distortions which provide measures for several order parameter fields. For instance, affine-type transforms of the local atomic lattice (i.e. expansions, compressions, and shears) serve as proxies for strain fields.[49-53] Additional examples include scalar lattice measurements such as oxygen octahedral tilt [54] or tetragonality[47, 49, 51, 54, 55], and vector quantities such as inversion-symmetry breaking from electrical polarization order parameter fields in displacive ferroelectrics.[49, 55-57]

While these techniques have been primarily used for qualitative studies, the information can be used to infer quantitative physics of materials via mesoscopic[58, 59] or discrete models.[60, 61] However, this approach further opens a question as to the role of non-observed degrees of freedom or unknown physics of interactions between the observed units, which can act as confounders or latent variables for observed behaviors. As an example, observation of the larger than expected Pt-Pt distance by Sohlberg[62] was interpreted as a presence of a capping OH group in a Pt trimer, providing insight into its catalytic activity. In this case, the discrepancy between theoretical prediction and quantitative experimental measurements was used to infer the presence of (an unobservable) structural element. A more complex case is the transition from the electronic carrier screening to oxygen vacancy screening at ferroelectric interfaces.[63] Another example is the observation of a small region of atomic scale distortion (similar to nanoislands) in a cleaved manganite crystal in an otherwise featureless scanning tunneling microscopy image, which was interpreted as a real-space image of a trapped polaron.[64]

However, these and many other examples rely on theoretical models that can fully explain the observed phenomena, and often require lengthy experiment and hypothesis testing cycles proposing and ruling out possible explanations for observed behavior. Often this approach requires development of novel theoretical models or introduces difficult to access approximations. In some cases, the direct information on the latent and confounding factors (beyond their existence) are unavailable, leading to the considerable uncertainty in interpretation of experimental results, as exemplified by multiple reports of ferroelectricity in ultrathin oxides that can be attributed both to ferroelectricity[65, 66] and electrochemical effects,[67] or the interplay between the two.[68-70] Generally, the discrepancy between the theory and observations (Bayesian surprise[71]) is used to guide scientific research via proposing the existence of new entities or new physical mechanisms.

Here, we explore whether predictability and uncertainty of physical behaviors can be used to gain insight into associated physical mechanisms in the presence of potential confounding, bias, and latent factors, via machine learning methods. We argue that experimentally observed correlations active over the large volumes of experimental data can be used to make a general prediction, whereas the regions where these correlative relationships are violated suggest the presence of new physics, either in the form of new laws or presence of latent variables as illustrated in Figure 1. For example, for an ideal ferroelectric material the local polarization can be expected to be strongly affected by the polarization in neighboring unit cells, but only weakly sensitive to the chemical composition. At the same time, in the vicinity of the phase boundaries (in composition space) or domain boundaries (in real space), the effects of the compositional fluctuations will be more pronounced, resulting in larger uncertainties in prediction of relevant properties. At the same time, the functional form of these relationships is generally unknown. Thus, an ability to correlate the descriptors to the observations with a universal function approximator that includes uncertainty quantification will enable determination of when such descriptors are adequate (and therefore, that our inclinations are supported), and when they appear inadequate to the task of functionality prediction. The latter can then lead directly to a search for new physical descriptors and further our understanding of these systems. This approach complements the classical physical paradigm of comparison to the known functional laws, or numerically derived models. Here, we use the Gaussian Processes as universal interpolators that allow uncertainty quantification in predictions via the posterior predictive density and explore the relationship between the predictability and *a priori* physical mechanisms.

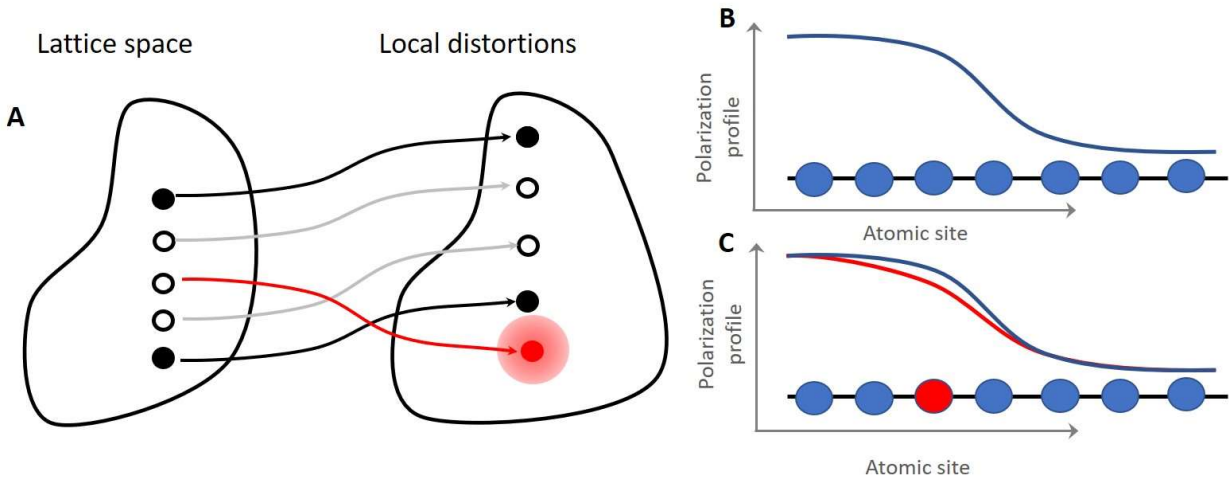

**Figure 1.** A Gaussian Process (GP) regression can, similar to a physical law with known functional form, be used to establish relationship between two sets of parameters. In this case, between the atomic coordinate within a lattice and the local polarization. In classical case, the relationship between the two is established based on known physical model, e.g. polarization profile across the domain wall. In GP case, the relationship is established based on multiple observations of the system providing interpolation function. Note that important aspect of GP approach is that prediction also yields uncertainty, allowing us to establish the significance of prediction and providing an independent channel of information on the process. Deviation from this behavior in certain locations can indicate the presence of hidden variables. For example, the presence of a dopant atom at a ferroelectric wall (B,C) will result in larger uncertainty in prediction of polarization at the certain lattice sites from that of the neighbors.

As a model system, we choose paradigmatic multiferroic material $BiFeO_3$ (BFO) with Sm substitutional doping of Bi grown as a continuous composition spread library. Pure BFO is a rhombohedral ferroelectric (R3c) with the polarization oriented along the <111> axis of a prototypical cubic unit cell. Upon Sm substitution of ~14%, the material transforms into an orthorhombic non-ferroelectric phase.[72, 73] The two are separated by the morphotropic phase boundary between symmetry-incompatible phases.[74, 75] The material structure and behavior in the vicinity of the MPB is highly complex, with many functional properties showing pronounced maxima and microstructure exhibiting the presence of the complex multiscale domain patterns, nanoscale regions, etc[76, 77]. While macroscopically many of these phases identify as monoclinic or triclinic, in the proximity to the MPB, while the corresponding atomistic structures remain unknown. Notably, the structure and functionality of materials at an MPB remain one of the most challenging questions in the physics of ferroic systems due to the multiple interactions between phases, domain structures, and disorder.

A continuous composition spread of $Bi_{1-x}Sm_xFeO_3$ in the range of $0 \leq x \leq 0.2$ was fabricated on a $SrTiO_3$ (001) substrate by pulsed laser deposition. A uniform gradient was produced by alternate ablation of $BiFeO_3$ and $SmFeO_3$ targets with a moving shadow mask. A chemical composition range, from 0 to 20% Sm was determined by wavelength dispersive spectroscopy (WDS) measurements. This composition range has been structurally characterized by X-ray diffraction and Piezoresponse Force Microscopy (PFM),[78, 79] and it displays a transition from highly ordered striped ferroelectric domains, to a ferroelectric mosaic, to a non-ferroelectric piezoresponse with increasing Sm%.[80]

Cross sectional atomic-resolution STEM imaging was performed for nominal $x = 0\%$, 7%, and 20% Sm concentrations by FIB lift out from a single composition-spread film. Data was collected along the [001]$_{pseudocubic}$ zone axis at 200kV on a Nion UltraSTEM with a High Angle Annular Dark Field (HAADF) detector, providing visualization of atomic-columns according to their mass-thickness. To preserve local spatial relationships against raster-scan artifacts the HAADF datasets were reconstructed from an orthogonal pair of HAADF images[81]. An example HAADF dataset for the 0% Sm $BiFeO_3$ endmember is shown in Fig. 2A,B. We quantify the local spatial relationships on the basis of atom positions, determined by fit as 2D Guassians, within a local neighborhood of the 5-cation perovskite unit cell as outlined in Ref [82]. Pertinent to the subsequent analysis we define descriptors for each B-site (Fe) centered unit cell which contain structural $\{V, \theta, \boldsymbol{a}, \boldsymbol{b}\}$, chemical $\{I_1, I_5\}$, and polarization $\{\boldsymbol{P}\}$ informational content, as defined in Table 1. They are, respectively, the unit cell volume (V), internal angle ($\theta$), in-plane lattice vector ($\boldsymbol{a}$), out-of-plane lattice vector ($\boldsymbol{b}$), mean HAADF intensity ($I_1$), A- to B-site intensity difference ($I_5$), and the vector of Fe-cation displacement from the A-site defined centro-symmetric positions ($\boldsymbol{P}$) which we hereafter refer to as the "polar displacement vector." Diagrams of the $V, \theta, \boldsymbol{a}, \boldsymbol{b}$, and $\boldsymbol{P}$ descriptors, derived from atom positions, are shown for an example unit cell in Figure 2B. Native units for these descriptors are pixels (datasets have a uniform scan resolution), with a real space conversion of 0.156 Å/pixel. $I_1$, and $I_5$ derive from the HAADF atomic-column intensities measured as the Gaussian weighted 9-pixel intensity centered at the atom fit position. These two selected HAADF intensity descriptors correspond to the average ($I_1$) and A- to B- site difference ($I_5$) of cation mass-thickness, and are indicators of sample thickness and cation composition variation, respectively.

| Sym. | U.C. Parameter Descriptions | Calculation |
| --- | --- | --- |
| **a** | In-plane lattice vector | $\frac{(A_{2,xy} - A_{1,xy}) + (A_{3,xy} - A_{4,xy})}{2}$ |
| **b** | Out-of-plane lattice vector | $\frac{(A_{4,xy} - A_{1,xy}) + (A_{3,xy} - A_{2,xy})}{2}$ |
| $\theta$ | Internal angle | Angle between **a**, **b** |
| $V$ | Internal volume | Volume of convex hull $A_{1,xy}$, $A_{2,xy}$, $A_{3,xy}$, and $A_{4,xy}$ |
| $I_1$ | Mean cation intensity | $\frac{A_1 + A_2 + A_3 + A_4 + B_1}{5}$ |
| $I_5$ | Cation intensity asymmetry A-site vs. B-site | $\frac{(A_1 + A_2 + A_3 + A_4)}{4} - B_1$ |
| **P** | Polar displacement vector | $\frac{A_{1,xy} + A_{2,xy} + A_{3,xy} + A_{4,xy}}{4} - B_{1,xy}$ |

**Table 1 – Unit cell descriptors**. Calculations correspond to A- and B-site 2D Gaussian atom fits per Fe-centered HAADF unit cell according to site labels in Fig 2B.

To explore the predictability of specific physical behaviors, we adopt the Gaussian Process method.[83-86] Gaussian process (GP) regression learns a function *f* over all source-target pairs $D = \{(x_1, y_1), \ldots (x_N, y_N)\}$, with each pair related by $y = f(x) + \varepsilon$, where $\varepsilon$ is Gaussian observation noise, by performing Bayesian inference in a function space, assuming that function *f* has a prior distribution $f \sim \mathcal{GP}(0, K_f(x, x'))$, where $K_f$ is a covariance function (kernel).[83] Naturally, GP is used as a powerful interpolator, where the covariance matrix of the GP posterior distribution serves as an estimator of the uncertainty in the interpolation. It should be noted that GP is a 'lazy' machine learning algorithm that is evaluated at runtime and does not require specific 'training'. The optimization of the GP entails maximizing the log marginal likelihood, by means of tuning the hyperparameters $\theta$ of the kernel function $K$, i.e.:

$$\log p(f(x)|\theta) = -\frac{1}{2} f(x)^T K(\theta, x, x')^{-1} f(x') - \frac{1}{2} \log \det(K(\theta, x, x')) - \frac{n}{2} \log 2\pi$$

In the above equation, we note that the first term is essentially a penalty term for failure to correctly predict function values, whereas the second term penalizes the model complexity. Here we used radial basis function kernel, $K(x, x') = \sigma^2 \exp\left(-\frac{\|x-x'\|^2}{2l^2}\right)$, where the variance $\sigma$ and the length scale $l$ are kernel hyperparameters, which are learned from the data by maximizing the log-marginal likelihood. For large datasets ($> 10^3$) computing the log marginal likelihood becomes nearly intractable and instead, a small set of $m$ function points is used as support or inducing variables, which are inferred along with kernel hyperparameters.[87] Here, we adapted the inducing points-based sparse implementation of GP from the *Pyro* probabilistic programming library,[88] which allows training a GP model with modern GPU accelerators, reducing the analysis time for datasets with $> 10^4$ points from hours to minutes. Once the model is trained, we can calculate the mean prediction and the associated variance for each point (see the accompanying notebook at https://git.io/Jv1Bn for the full code of GP regression). Here, the high variance reflects higher uncertainty in GP prediction.

To gain insight into the predictability of the physical parameters in these materials, we first explore the polarization behavior in the system. Here, we assume that the material possesses an electrical polarization field with a proxy representation by the polar displacement vector descriptor ***P***, defined as the off-centering of the B-site cation. The following treatment is on the two components of ***P***, noted as $P_x(i,j)$ and $P_y(i,j)$, where $i$ and $j$ are the lattice position of the unit cell. We explore whether the polarization at the individual lattice site can be predicted from the polarization in the surrounding area, i.e. regressions

$$GP_{ny \to y}(P_y(i \pm 1, j \pm 1)) \to P_y(i,j) \tag{1}$$

between an 8-component vector of polarization values in the 8 adjacent lattice sites and polarization in the given site. We further explore the predictability of the polarization component from the other polarization component, i.e. regression

$$GP_{nx \to y}(P_x(i \pm 1, j \pm 1)) \to P_y(i,j) \tag{2}$$

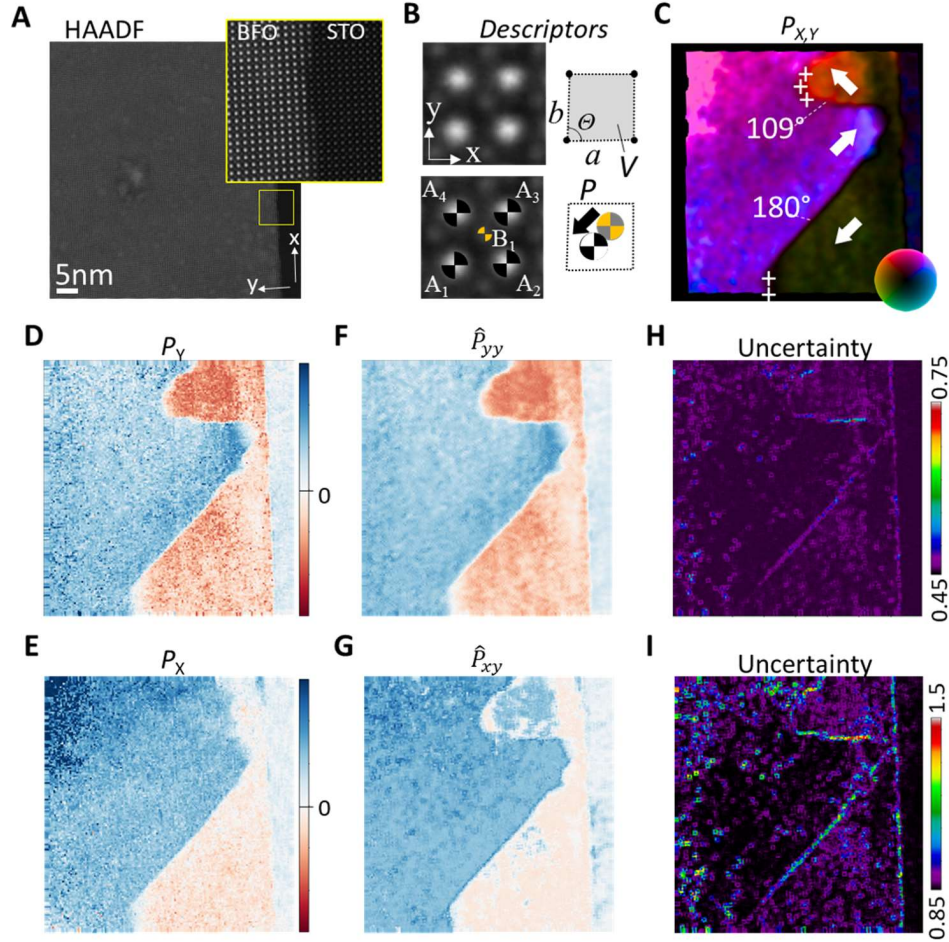

**Figure 2:** STEM data and GP predictions for $x = 0$, $Bi_{1-x}Sm_xFeO_3$ endmember. **(A)** HAADF STEM dataset for the $BiFeO_3$ film, $SrTiO_3$ substrate at right, with highlighted region magnified in the inset. **(B)** An example single Fe-centered unit cell from A, shown rotated with corresponding cation position and labels. Diagrams for spatially determined descriptors are shown at right. **(C)** Colorized map of the polar displacement vector descriptor, ***P***, depicting the polydomain ferroelectric structure with directions and domain walls types labeled. **(D, E)** The experimental y- and x-components of ***P***. **(F, G)** Corresponding $\hat{P}_{yy}$ and $\hat{P}_{xy}$ GP predictions from 8-neighbor $P_y$ values with **(H, I)** uncertainty maps. All units are in pixels, with real space conversion of 0.156 Å/pixel. The outer radius of the ***P*** color legend **(C)** and the component colorscales **(D-G)** are ± 5.97 pixels.

The spatial degeneracy of the ferroelectric polarization frequently results in formation of polydomain configurations to minimize stray field energies and is a product of growth kinetics.

Such a polydomain structure is observed for the 0% Sm ferroelectric BiFeO$_3$ endmember shown in Fig. 2C-E. This image contains three domains indicated by directional arrow labels in Fig. 2C. The sharp boundaries, i.e. domain walls, are classified according to their polarization rotation across the wall. The domain walls present in the image are 109° and 180°, as labeled in Fig 2C, identified by their 2D projected rotations (where the z-axis rotation must be inferred) and conforming to known preferential (100) and (110) planes, respectively. In both instances the domain walls deviate from their charge-neutral planes to ones with divergent polarization orientations, labeled with "+" symbols representing their bound electrical charges, classified as charged domain walls.

Not surprisingly, the predicted polarization, denoted as $\hat{P}_{yy}$, from the $P_y$ polarization neighborhood shows values and distribution very close to the original values (Fig. 2F), with the visible effect of GP being reduction and smoothing of the noise, i.e. essentially a convolution effect. However, non-trivial behavior emerges in the uncertainty maps (Fig. 2G), where significant uncertainty of prediction is observed in the vicinity of domain walls and in a number of distributed locations throughout the material. Notably, the prediction uncertainty is zero inside the substrate, i.e. local laws are well-defined.

The predicted polarization $\hat{P}_{xy}$ from the $P_y$ polarization neighborhood as compared to the true polarization distribution (Fig. 2G) shows considerably more interesting behavior. Note that in most locations of the image, the predicted polarization is close to the true polarization, as can be expected since GP is local interpolator. The prediction is reliable in the vicinity of the 109º domain wall. However, in the vicinity of the charged domain wall the polarization shows strong deviations from the true polarization distribution, with clearly visible regions of opposite polarization direction. Careful examination of all polarization components shows that there may be fluctuations in the true polarization field; however, these are relatively minor. Therefore, the predicted "anomalous" polarization in the vicinity of the charged DW may be an indicator of confounding physical factors.

Interestingly, the uncertainty map in Fig. 2I shows overall higher uncertainty in prediction compared to $GP_{ny \to y}$, with the larger uncertainty values in the vicinity of DWs and substrate interface. However, the prediction uncertainty has spatial distribution generally different from the difference between predicted and observed polarization.

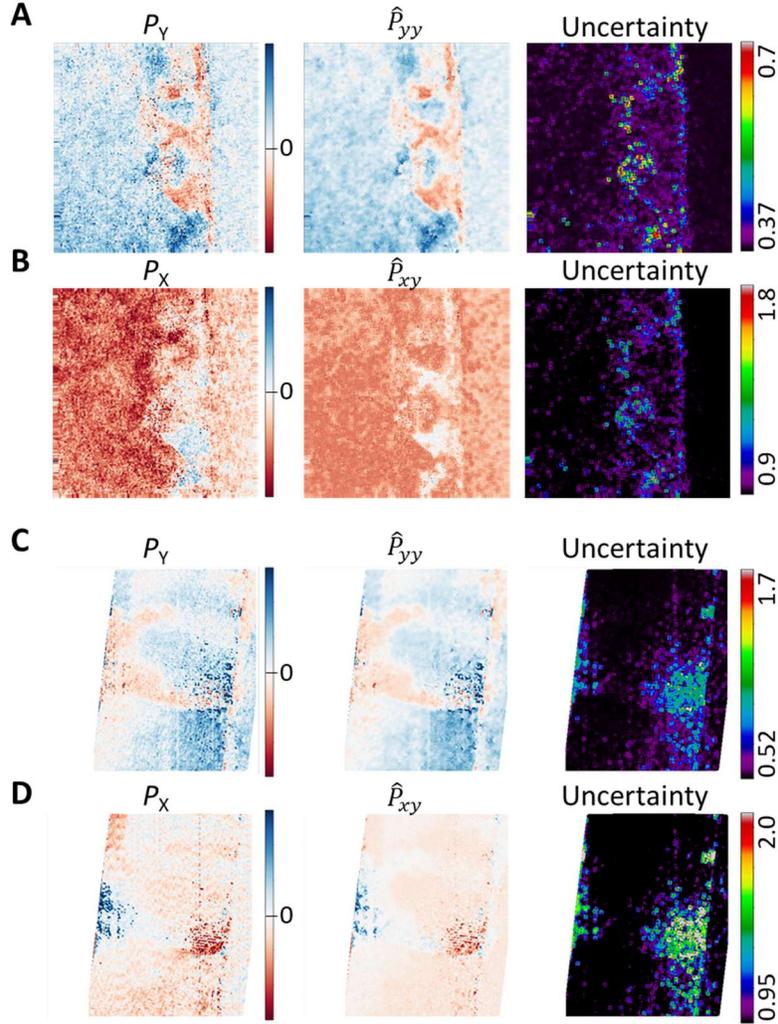

**Figure 3:** $P_x$, $P_y$ data and $P_y$ GP predictions for x=0.07 **(A,B)** and x=0.2 **(C,D)** $Sm_xBi_{1-x}FeO_3$ film compositions. **(A)** Original $P_y$ data and $\hat{P}_{yy}$ GP prediction for $x = 0.07$ and **(B)** $P_x$ data and $\hat{P}_{xy}$ GP prediction. **(C)** Original $P_y$ data and $\hat{P}_{yy}$ GP prediction for $x = 0.2$ and **(D)** $P_x$ data and $\hat{P}_{xy}$ GP prediction. All units are in pixels, with 0.156 Å/pixel, ***P*** component color scales are ± 5.97.

Similar analyses for the other two compositions corresponding to the intermediate phase ($x = 0.07$) and the orthorhombic phase ($x = 0.2$) are shown in Fig. 3. Here, the intermediate phase (Fig. 3 A,B) shows the presence of a ferroelectric phase with non-zero polarization values in the vicinity of the interface and a disordered phase in the film. The nature of this disordered phase is unclear, and can correspond to a nanodomain system averaged in the z-axis direction of the electron beam, or a true disordered intermediate phase. Similar to the rhombohedral $x = 0$ sample, the $\hat{P}_{yy}$ image shows features similar to the true image $P_y$ (Fig 3A), albeit with considerably lower noise level and a slight blur. The corresponding uncertainty map shows regions with high uncertainty at the boundaries between the ferroelectric and disordered phase. Interestingly these features seem to be new, and while they partially overlap with the domain walls, there is not a one to one correspondence. Again, the uncertainty for prediction within the substrate is uniform and small. The true $P_x$ image, predicted $\hat{P}_{xy}$, and corresponding uncertainty are shown in Figure 3B. Notably, the uncertainty in GP is very similar for $\hat{P}_{yy}$ and $\hat{P}_{xy}$.

Finally, similar analysis is performed for the $x = 0.2$ orthorhombic phase (Fig 3 C,D). In this case, the polarization distribution show only weak contrast within the image and demonstrates small inclusions with period doubling (discussed later). For this composition, both $\hat{P}_{yy}$ and $\hat{P}_{xy}$ are very close to the true distributions $P_y$ and $P_x$, and corresponding uncertainty maps show increase of uncertainty in the vicinity of the modulated phase regions.

To summarize, for all compositions explored, the predicted and uncertainty maps demonstrate additional spatial features. Some of these are concentrated at the regions where polarization changes rapidly, including domain walls and the interface. Yet others are concentrated in the "cloud-like" regions within the material, and do not bear obvious resemblance with the individual atomic descriptor maps. Hence, we conclude that the prediction and uncertainty maps contain nontrivial information or at least highlight the discrepancies where locally observed polarization differs from that expected given the average polarization behavior in the sample, pointing to new physical phenomena or the presence of latent variables.

We further proceed to explore the predictability and prediction uncertainty between the structural descriptors and polarization. For this, we perform the regressions

$$GP_{s \to y}(\mathbf{S}_{ij}) \to P_y(i,j) \tag{3}$$

where $\mathbf{S}_{ij}$ is the parameter vector of local structural descriptors.

In principle, such analysis can be performed using all the permutations of the available structural descriptors and their gradients (e.g. bag of features[89] approach), etc. to obtain comprehensive insight into functional relationships between the descriptors. Given the computational constraints and being informed by the general physics of ferroic materials, here we explore five groups of descriptors, namely $\{V, \theta, \boldsymbol{a}, \boldsymbol{b}, I_1, I_5\}$, $\{\boldsymbol{a}, \boldsymbol{b}\}$, $\{V, \theta\}$, $\{V, \theta, \boldsymbol{\alpha}, \boldsymbol{b}\}$, and $\{I_1, I_5\}$. This choice of groups is made given that $I_1$, $I_5$ are parameters describing local chemical composition, $V$ is related to the local phase state and composition (via Vegard terms), θ is the measure of the unit cell deformation from cubic/tetragonal into modulate phase, *a* and *b* define tetragonality of the unit cell. These variables are not independent; however, they can be considered as semiquantitative descriptors for composition, ferroelectricity, and proximity to the orthorhombic phase, respectively. Notably, all these descriptors are invariant with respect to the $C_2$ rotations, whereas polarization components change sign. Hence, the regression Eq. (3) explores whether polarization can be predicted from this set of structural and chemical descriptors and provide the associated uncertainties.

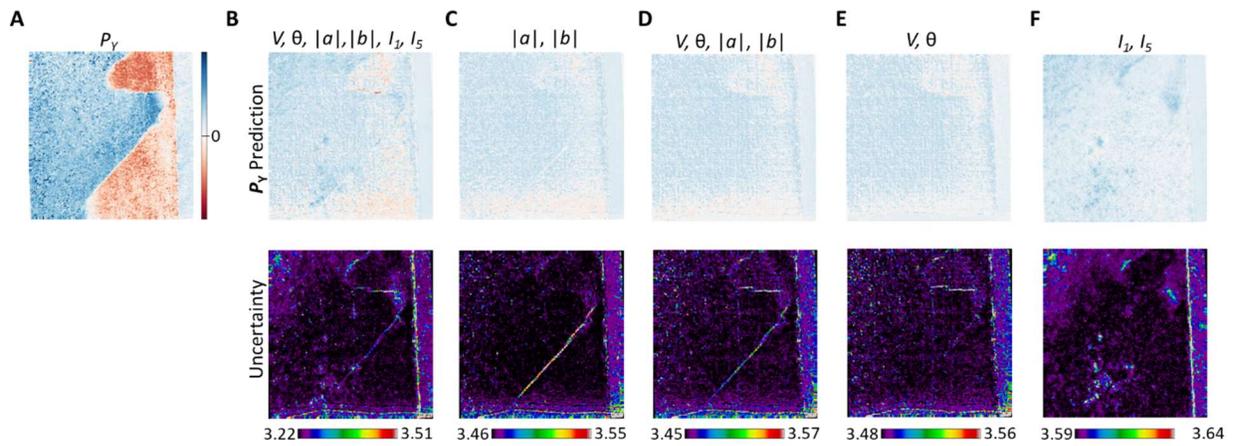

**Figure 4:** Original $P_y$ data and GP predictions $P_{ys}$ from structural descriptor for rhombohedral phase, $x = 0$. **(A)** original $P_y$ data. **(B-F)** GP predictions from structural descriptor sets: **(B)** $\{V, \theta, \alpha, b, I_1, I_5\}$, **(C)** $\{a, b\}$, **(D)** $\{V, \theta\}$, **(E)** $\{V, \theta, \alpha, b\}$, and **(F)** $\{I_1, I_5\}$. All $P_y$ colorscales are ± 5.97 pixels.

The analysis for rhombohedral BiFeO$_3$ is shown in Figure 4. Here, the top row visualizes the true polarization distribution $P_y$ (Fig 3A) and the predictions $\hat{P}_{ys}$ for the different groups of structural descriptors (Fig 3B-F). Upon examining the characteristic features in the images, we note that none of the groups allow for reliable reconstruction of polarization field, as can be expected from the physics of the problem. At the same time, the groups containing $\{V, \theta\}$, as descriptors allow to predict the changes in polarization in the charged domain and separate the ferroelectric phase and the substrate. The prediction based on $\{a, b\}$ is completely non-discriminative, and only the anomalies at the domain walls are visible. Finally, chemical descriptors $\{I_1, I_5\}$. yield the maps resembling chemical disorder in the system, but do not contain any elements of the true polarization distribution or phase contrast between ferroelectric and substrate.

In comparison, the uncertainty maps shown in the bottom row of Fig. 4 show considerably richer behavior. Here, the prediction uncertainties for descriptor groups containing $\{a, b\}$ show clear anomaly at the 109° wall, which is invisible for $\{V, \theta\}$ and $\{I_1, I_5\}$ descriptors. At the same time, chemical descriptors $\{I_1, I_5\}$ give rise to an uncertainty maximum at the ferroelectric-substrate interface. While this feature can be noticed in three other maps, it is much weaker. The charged domain wall is visible only in the maps containing $\{V, \theta\}$. Finally, $\{I_1, I_5\}$ seem to provide regions of higher uncertainty uncorrelated to the true potential distributions and potentially related to the presence of chemical disorder.

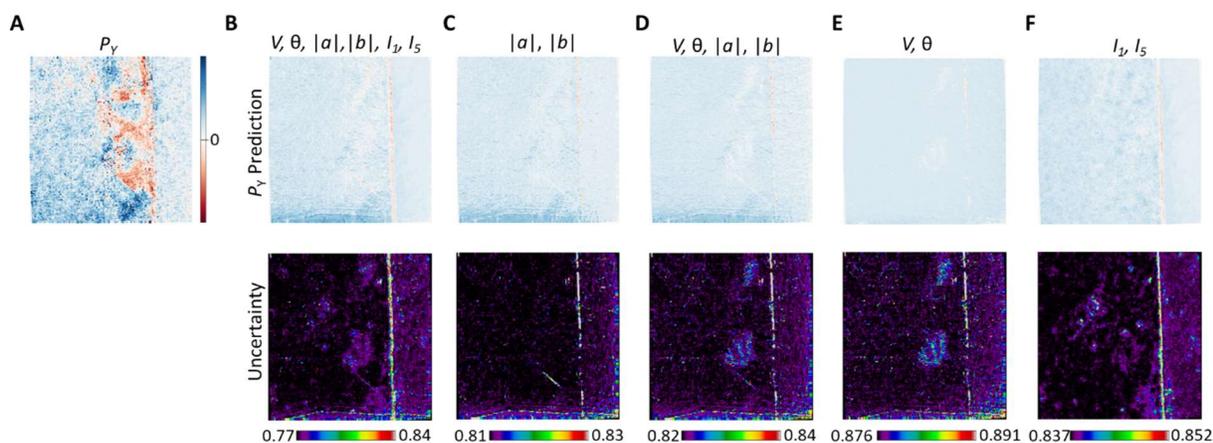

**Figure 5:** Original $P_y$ data and GP predictions $P_{ys}$ from structural descriptor for $x = 0.07$. **(A)** original $P_y$ data. **(B-F)** GP predictions from structural descriptor sets: **(B)** $\{V, \theta, \boldsymbol{a}, \boldsymbol{b}, I_1, I_5\}$, **(C)** $\{\boldsymbol{a}, \boldsymbol{b}\}$, **(D)** $\{V, \theta\}$, **(E)** $\{V, \theta, \boldsymbol{a}, \boldsymbol{b}\}$, and **(F)** $\{I_1, I_5\}$. All $P_y$ colorscales are ± 5.97 pixels.

A similar analysis for the intermediate composition, $x = 0.07$, is shown in Fig. 5. Similar to the ferroelectric phase, in general the predicted polarization field is very dissimilar from the true polarization, with only the maps containing $\{V, \theta\}$ group showing features resembling polarization field. At the same time, $\{\boldsymbol{a}, \boldsymbol{b}\}$ group clearly visualizes the 180° domain wall. Note the highly visible contrast in the predicted map and associated uncertainty map. In this case only one of the domain walls shows in this manner. Finally, $\{I_1, I_5\}$ and to some extent $\{V, \theta\}$ give rise to contrast at the Sm:BFO – substrate interface. The corresponding uncertainty maps show blob features primarily associated with $\{V, \theta\}$ component, interface uncertainty associated with $\{I_1, I_5\}$, and clearly visible 109° domain wall associated with the $\{a, b\}$.

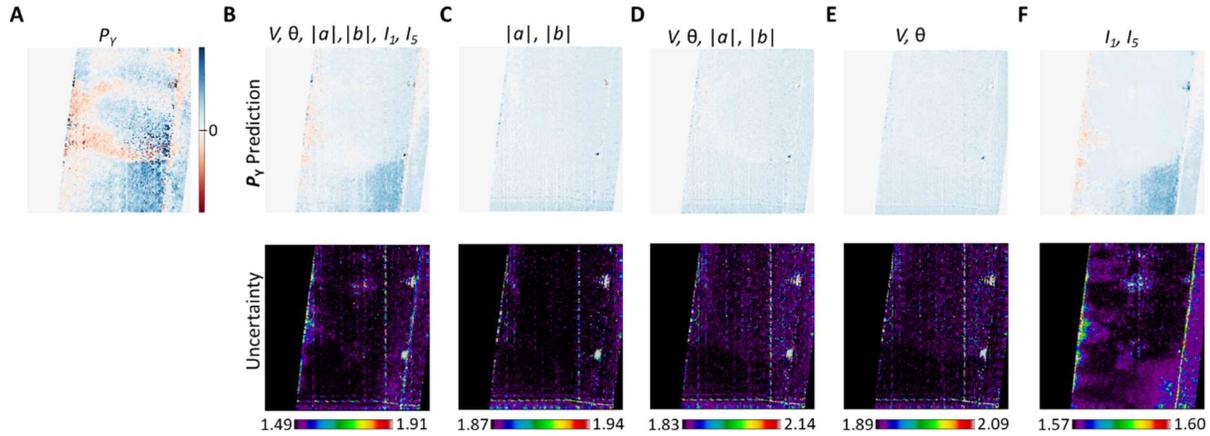

**Figure 6:** Original $P_y$ data and GP predictions $P_{ys}$ from structural descriptor for $x = 0.2$. **(A)** original $P_y$ data. **(B-F)** GP predictions from structural descriptor sets: **(B)** $\{V, \theta, \alpha, b, I_1, I_5\}$, **(C)** $\{a, b\}$, **(D)** $\{V, \theta\}$, **(E)** $\{V, \theta, \alpha, b\}$, and **(F)** $\{I_1, I_5\}$. All $P_y$ colorscales are ± 5.97 pixels.

Finally, this analysis for the high Sm concentration, $x = 0.2$, composition is shown in Fig. 6. Here, the examination of the predicted polarization fields and associated uncertainties reveals that $\{a, b\}$ group of variables does not allow prediction and associated uncertainty maps are essentially featureless. The $\{I_1, I_5\}$ group shows an interesting effect of image segmentation, with top part of the image being featureless and bottom showing some contrast. The associated uncertainty map shows uncertainties at the film-substrate interface. However, very interesting behavior is associated with the $\{V, \theta\}$ group of variables that clearly shows the emergence of the modulated phase throughout the top 2/3 of the image. This phase is essentially unnoticeable in the original polarization image. What is particularly noteworthy is that this predicted ordering differs from the original one in the data set.

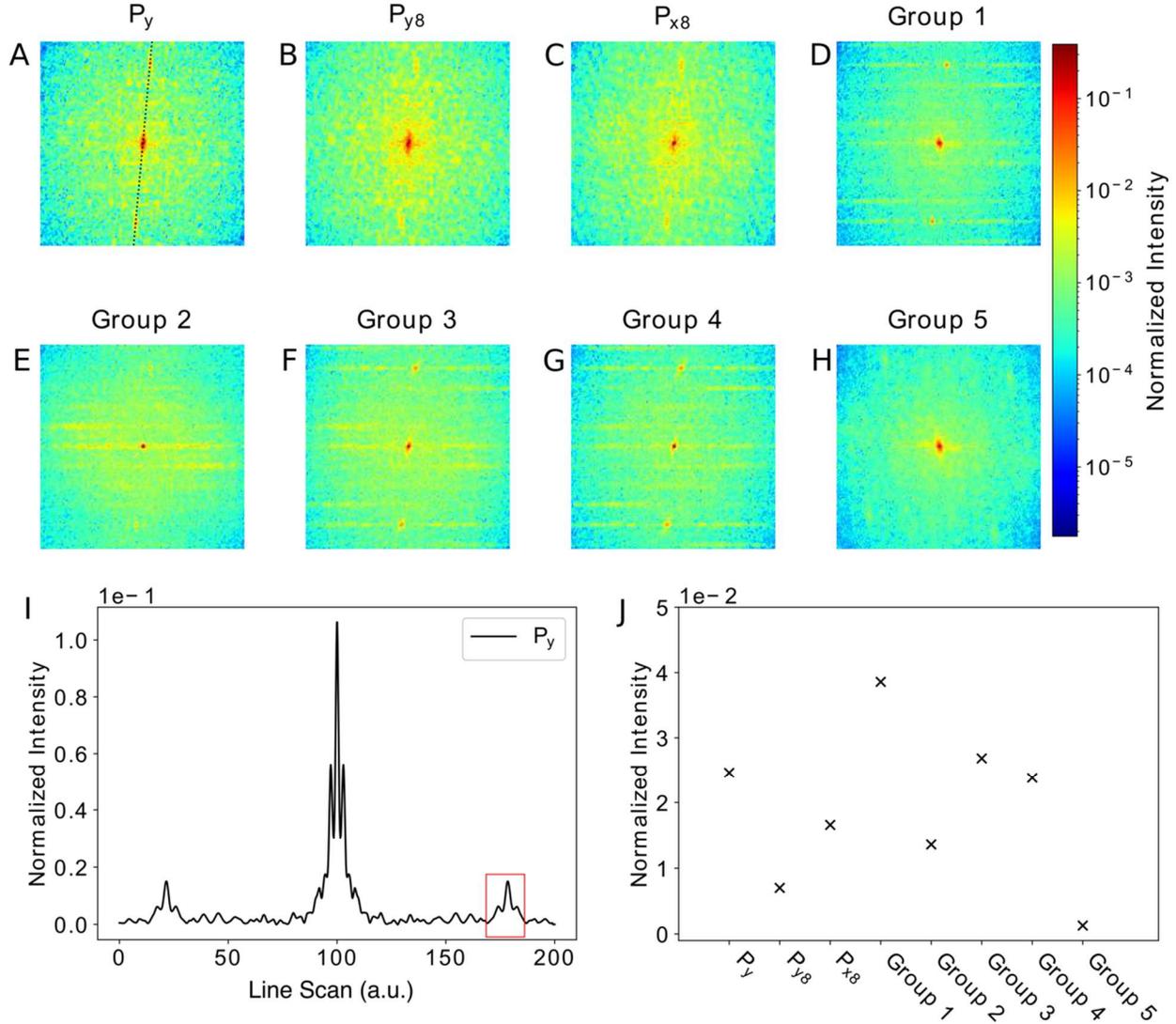

**Figure 7:** FFTs from $Bi_{0.8}Sm_{0.2}FeO_3$ of the **(A)** $P_y$ data, and $P_y$ predictions from **(B)** $P_{y8}$, **(C)** $P_{x8}$, and **(D-H)** prediction groups 1-5 consisting of $\{V, \theta, a, b, I_1, I_5\}$, $\{a, b\}$, $\{V, \theta, a, b\}$, $\{V, \theta\}$, and $\{I_1, I_5\}$. **(I)** Line profile of FFT of $P_y$ data going through the main diffraction spots, indicated by the black line in (A), and **(J)** the normalized FFT intensity of the diffraction spot, outlined in red in (I), as a function of prediction group. Note that (I) predicted FFT intensity in group 1 is twice higher than in original image, and group 5 predicts a different ordering then original data set, in <11> rather than <01> direction.

The Fourier transform of each predicted $P_y$ image can be seen in Figure 7 A-H, where each FFT was calculated with a Hanning window to account for edge effects. To account for differences in signal strength, each FFT is normalized to the area under the cross-section of the central spot. We can see that the maximum FFT intensity is observed in the original image, Figure 7A, with some superlattice spots appearing just above the background signal. Similar diffraction spots can be seen in many of the prediction FFTs including those predicted from $P_{y8}$ (the 8 neighbor $P_y$ values as used for the $\hat{P}_{yy}$ GP prediction), $P_{x8}$ (the 8 neighbor $P_x$ values), $\{V, \theta, \boldsymbol{a}, \boldsymbol{b}, I_1, I_5\}$, $\{V, \theta, \boldsymbol{a}, \boldsymbol{b}\}$, and $\{V, \theta\}$. In order to compare the intensities of the diffracted spots, line profiles of each FFT was taken such that the diffraction spots are included, as seen in Figure 7A,I. The normalized intensities of the diffracted spots for each descriptor group are shown in Figure 7J. The maximum intensity of the diffracted spots occurs when using the full set of descriptors, $\{V, \theta, \boldsymbol{a}, \boldsymbol{b}, I_1, I_5\}$, with dominant contributions coming from the $\{V, \theta, \boldsymbol{a}, \boldsymbol{b}\}$, and $\{V, \theta\}$. Here, further analysis established that $\theta$ is the preponderant origin of the modulation. In comparison, $\{I_1, I_5\}$ and $\{\boldsymbol{a}, \boldsymbol{b}\}$ correspond to almost zero predicted modulation.

While the diffracted spots appear in the FFT, the corresponding ordered region in the real space ground truth ($\boldsymbol{P_y}$) image is difficult to identify, though they can clearly be seen in the real space images of the predictions from $\{V, \theta, \boldsymbol{a}, \boldsymbol{b}, I_1, I_5\}$, $\{V, \theta, \boldsymbol{a}, \boldsymbol{b}\}$, and $\{V, \theta\}$. Completely unexpectedly, the FFT for $\{I_1, I_5\}$ shows a modulation of a different type, corresponding to (½, ½), however they are very weak. The detailed analysis of the data suggests that these might be originating from the bottom right corner of the sample, as can be confirmed by inverse FFT analysis. Note that the prediction of the period doubling phase by GP is a highly non-trivial result. We pose that a possible explanation for this behavior is that the compositional and structural descriptors are that of the orthorhombic phase, but its emergence is suppressed due to the confinement and mechanical depolarization effects.

To explore whether behaviors and uncertainties derived using Gaussian Process regression can be derived via linear analysis, we explored the linear correlation structure between the predictors, predictions, and the ground truth images using canonical correlation analysis (CCA).[90] In this analysis, two sets of multi-dimensional variables (X, Y) are connected by finding linear combinations of variables which maximize their correlation. New variables are defined by the linear combinations with weights $\mathbf{w_a}$ and $\mathbf{w_b}$ such that $U = \mathbf{w_a'X}$ and $V = \mathbf{w_b'Y}$. The weights ($\mathbf{w_a}$, $\mathbf{w_b}$) are determined by maximizing Equation 4, subject to unit variance, $\mathbf{w_a'X w_a} = \mathbf{w_b'Y w_b} = 1$, to account for scaling:

$$Cor(X,Y) = \frac{Cov(U,V)}{[\sqrt{Var(U)})\sqrt{Var(V)}]} \tag{4}$$

The Pearson correlation coefficient, which quantifies the magnitude of association between two variables,[91] is then calculated between the learned pair of canonical variables. Specifically, the correlation coefficients are calculated given the (X, Y) pairs: ({descriptor group}, $P_y$), ({predictions from descriptor group}, $P_y$), and ({descriptor group},{predictions from descriptor group}), where ({ }, $P_y$) is each descriptor group or prediction from a descriptor group. In the last case, each descriptor group is only compared to the prediction derived from that group, i.e. no cross-comparisons are made between descriptor groups and predictions.

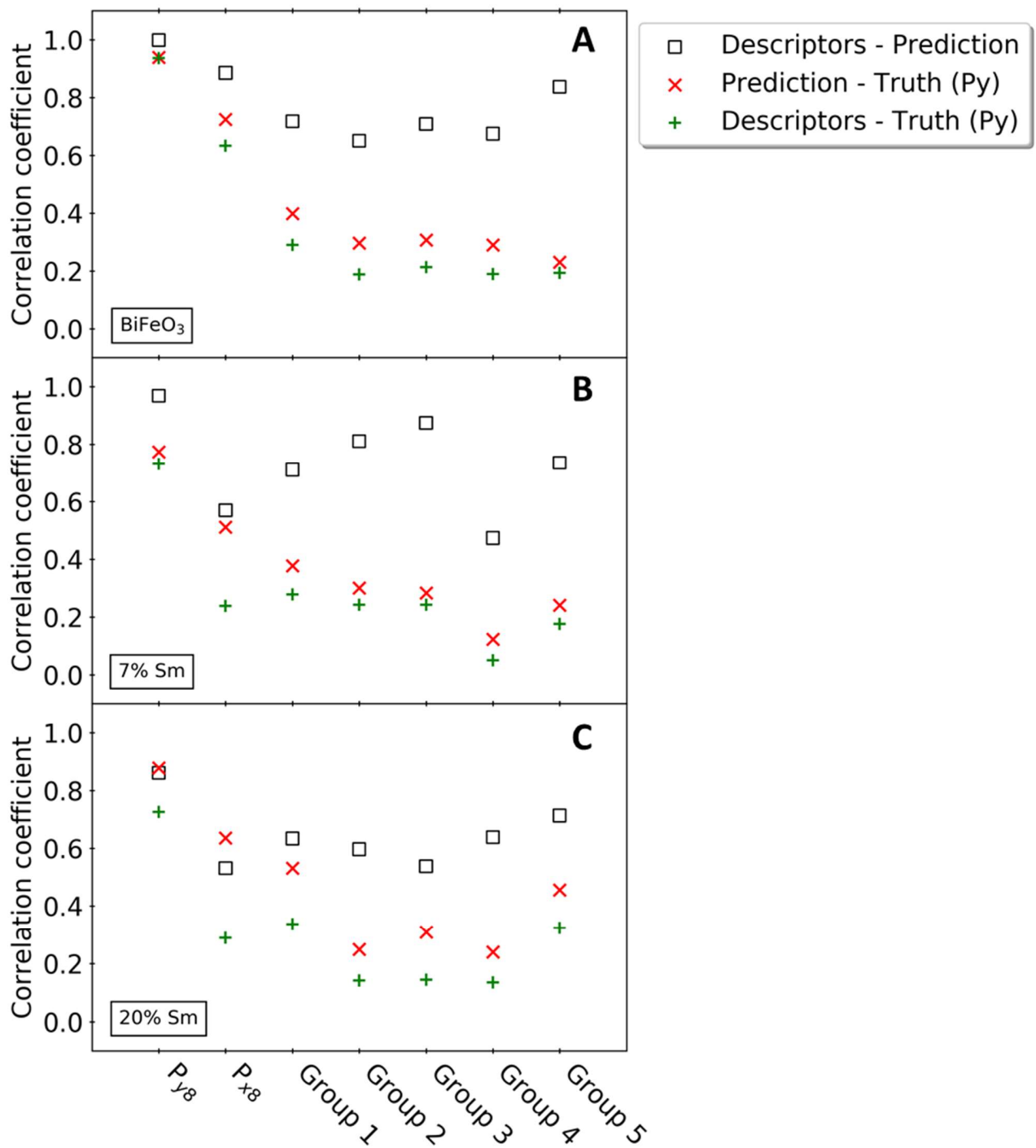

**Figure 8**. Correlation coefficients for the three compositions **(A)** $BiFeO_3$, **(B)** $Bi_{0.93}Sm_{0.07}FeO_3$, and **(C)** $Bi_{0.8}Sm_{0.2}FeO_3$ where Group 1-5 are $\{V, \theta, \boldsymbol{a}, \boldsymbol{b}, I_1, I_5\}$, $\{\boldsymbol{a}, \boldsymbol{b}\}$, $\{V, \theta, \boldsymbol{a}, \boldsymbol{b}\}$, $\{V, \theta\}$, and $\{I_1, I_5\}$ respectively.

The CCA analysis of the pairwise combinations between the ground truth, descriptors, and GP predictions is shown in Figure 8 for the chosen groups of descriptors and for three studied compositions. As a general trend, the descriptors show relatively high correlations with the prediction, which is maximal for the pure $BiFeO_3$ and decreases for higher Sm concentrations, across all descriptor groups. However, this correlation is well below unity and can be as small as ~0.5 for high Sm content BFO. It is also important to note that the linear correlation explored here provides average information across image. Hence, the behaviors observed at the localized structural elements (domain walls, interfaces, etc.) has to be interpreted independently, as above.

The correlation between the prediction and truth is generally above the correlation between the descriptors and predictors, clearly illustrating that non-linear GP regression is capable of better prediction than linear estimates. This improvement is particularly pronounced for the high Sm composition, where corresponding correlation coefficients are higher by as much as factor of 2.

The analysis of the quality of prediction (i.e. correlation between prediction and truth) for different descriptor groups across the compositions reveals several universal observations. In all cases, it appears that the $P_{y8}$ group predicts $P_y$ extremely well. This behavior is expected, given the strong correlations between the polarization for adjacent unit cells. The correlation between $P_{x8}$ and $P_y$ is fairly strong in BFO and becomes considerably weaker for the high Sm composition. Finally, extremely interesting behaviors are observed for the structural and chemical descriptors. Here, for the strongly ferroelectric material the best prediction is achieved with the structural descriptors including $V$, $\theta$, *a*, and *b*, whereas inclusion of chemical descriptors $I_1$, $I_5$ considerably lowers the prediction. The opposite behavior is observed for the intermediate and high Sm concentrations, where the quality of prediction improves considerably once the chemical effects are included.

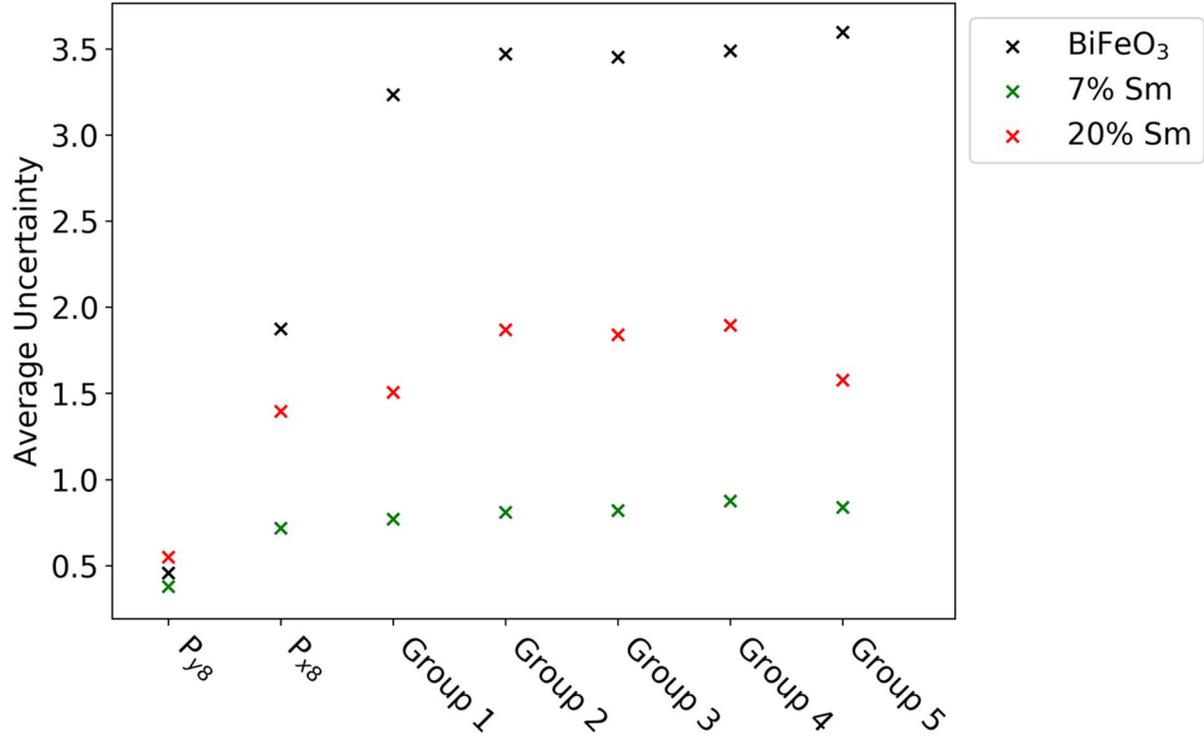

**Figure 9.** Average uncertainty in prediction for all three compositions and each descriptor group where Group 1-5 are $\{V, \theta, \boldsymbol{a}, \boldsymbol{b}, I_1, I_5\}$, $\{\boldsymbol{a}, \boldsymbol{b}\}$, $\{V, \theta, \boldsymbol{a}, \boldsymbol{b}\}$, $\{V, \theta\}$, and $\{I_1, I_5\}$ respectively.

The behavior of GP prediction uncertainty across the compositions and descriptor groups is shown in Fig. 9. Note that uncertainty is available only for the Gaussian (or more generally, Bayesian) methods. As expected, for polarization descriptors $P_{y8} \rightarrow P_y$ the uncertainty is lowest, i.e. prediction is highly reliable.

At the same time, highly non-trivial behavior is observed for the predictions based on the different polarization component $P_{x8}$, and structural and chemical descriptors. Here, the uncertainty in prediction is highest for BFO. This makes sense, since in strongly ferroelectric BFO polarization is less sensitive to the chemical and structural factors since the corresponding potential well is deepest and depolarization effects are strongest. The orthorhombic phase shows the intermediate uncertainty, presumably due to the fact that polarization is close to zero. Interestingly, the lowest uncertainty in prediction is observed for intermediate phase. We ascribe this behavior to the fact that it is a mixture of ferroelectric and non-ferroelectric phases which have different lattice parameters, and ferroelectric phase is polarized in one direction. Hence polarization is most sensitive to the local compositional fluctuations, allowing for good predictability. Interestingly, this lability is also the origin of the high functional responses of morphotropic materials. At the same time, this information is absent in a linear correlation analysis.

Similar to correlation analysis, including of additional parameters slightly reduces the uncertainty, e.g. prediction from $\{V, \theta, \boldsymbol{a}, \boldsymbol{b}\}$ reduces uncertainty compared to $\{\boldsymbol{a}, \boldsymbol{b}\}$ and $\{V, \theta\}$. Inclusion of $I_1$, $I_5$ considerably reduces uncertainty for intermediate, somewhat less high Sm material, but not pure one. Again, this behavior comports well with the expectations, since compositional effects can be especially pronounced for regions with phase competition.

To summarize, here we explored whether the predictability and quantified uncertainty of physical behaviors can be used to determined manifest and latent physical mechanisms. Classically, this approach is implemented via e.g. comparison of theory to experiment, to establish the presence of latent physical variables or unknown mechanisms, and subsequently refine their functional form and associated model parameters. Here, we implement this paradigm via analysis of predictability and uncertainty of Gaussian process regression between multiple spatially distributed degrees of freedoms. In other words, we explored whether local deviations of the characteristic behaviors established globally can be used to identify the latent local physical behaviors.

This approach is implemented for multiferroic Sm-doped $BiFeO_3$ across the broad composition range as a model. Using the atomic resolution Scanning Transmission Electron Microscopy (STEM) data, we directly visualize the atomic structure and structural, physical, and chemical order parameter fields for the material. GP regression is used to establish the predictability of local polarization field from the different groups of parameters, including the adjacent polarization values and several combinations of physical and chemical descriptors including lattice parameters, column intensities, etc. We observe that certain elements of materials domain structure including charged and uncharged domain walls and interface with substrate are best predicted with certain specific combinations of descriptors, and this predictability and associated uncertainties are consistent across the composition series. For example, the 180 domain walls are best predicted from unit cell sizes, whereas charged walls are best predicted from the molar volume and angles. Interestingly the GP regression is shown to considerably exceed predictions based on linear correlative models. The associated physical mechanisms generally agree with the expected behaviors for ferroelectric materials across the morphotropic phase boundary with the non-ferroelectric phase. Remarkably, we show that uncertainty in prediction provides a new channel of information, presumably allowing to identify the regions with latent mechanism, in this case compositional fluctuations. Finally, and surprisingly the GP regression predicts the experimentally unobserved ordering in the Sm-doped BFO phase.

Overall, we believe that proposed approach can be broadly used in exploration of potential novel physical mechanisms in both experimental and modeling datasets. For instance, such tools may be extremely useful for interpretation of molecular dynamics data of disordered systems, or for analysis of nanoscale segregation in alloys by inspection of data form atom probe tomography. Future work can focus on more advanced functional interpolators, such as neural network-induced Gaussian Process models, or Bayesian neural nets, as well as more complex descriptors that take into consideration longer range effects (e.g., graph-based methods). Similarly, one may apply this methodology to determining mechanisms in light-induced dynamics with local probes – for instance, exploration of photoconductive effects and ionic transport in perovskite photovoltaics, or for isolating factors involved in ferroelectric fatigue from nanoscale imaging and chemical spectroscopy.

**Acknowledgements:**

This effort (electron microscopy, feature extraction) is based upon work supported by the U.S. Department of Energy (DOE), Office of Science, Basic Energy Sciences (BES), Materials Sciences and Engineering Division (S.V.K., C.N.) and was performed and partially supported (MZ, RKV) at the Oak Ridge National Laboratory's Center for Nanophase Materials Sciences (CNMS), a U.S. Department of Energy, Office of Science User Facility. The work at the University of Maryland was supported in part by the National Institute of Standards and Technology Cooperative Agreement 70NANB17H301 and the Center for Spintronic Materials in Advanced infoRmation Technologies (SMART) one of centers in nCORE, a Semiconductor Research Corporation (SRC) program sponsored by NSF and NIST. A.N.M. work was partially supported by the European Union's Horizon 2020 research and innovation program under the Marie Skłodowska-Curie (grant agreement No 778070). The work at North Carolina State University was supported in part by the National Science Foundation Grant No. DGE-1633587. One of the authors (SVK) gratefully acknowledges inspiring discussions with Vint Cerf (Google) and Judea Pearl (UCLA) that stimulated this direction of research.

**Data Availability:**

The code and data are available without restrictions at https://git.io/Jv1Bn .

# References

1. I. Grinberg, M. R. Suchomel, P. K. Davies and A. M. Rappe, Journal of Applied Physics **98** (9) (2005).
2. D. Damjanovic, Rep. Prog. Phys. **61** (9), 1267-1324 (1998).
3. M. A. Tanatar, N. Spyrison, K. Cho, E. C. Blomberg, G. T. Tan, P. C. Dai, C. L. Zhang and R. Prozorov, Phys. Rev. B **85** (1) (2012).
4. S. J. Kuhn, M. K. Kidder, D. S. Parker, C. dela Cruz, M. A. McGuire, W. M. Chance, L. Li, L. Debeer-Schmitt, J. Ermentrout, K. C. Littrell, M. R. Eskildsen and A. S. Sefat, Physica C-Superconductivity and Its Applications **534**, 29-36 (2017).
5. Y. Zheng, P. M. Tam, J. Q. Hou, A. E. Bohmer, T. Wolf, C. Meingast and R. Lortz, Phys. Rev. B **93** (10) (2016).
6. H. Xiao, T. Hu, S. K. He, B. Shen, W. J. Zhang, B. Xu, K. F. He, J. Han, Y. P. Singh, H. H. Wen, X. G. Qiu, C. Panagopoulos and C. C. Almasan, Phys. Rev. B **86** (6) (2012).
7. H. Xiao, T. Hu, A. P. Dioguardi, N. apRoberts-Warren, A. C. Shockley, J. Crocker, D. M. Nisson, Z. Viskadourakis, X. Tee, I. Radulov, C. C. Almasan, N. J. Curro and C. Panagopoulos, Phys. Rev. B **85** (2) (2012).
8. J. Chakhalian, J. M. Rondinelli, J. Liu, B. A. Gray, M. Kareev, E. J. Moon, N. Prasai, J. L. Cohn, M. Varela, I. C. Tung, M. J. Bedzyk, S. G. Altendorf, F. Strigari, B. Dabrowski, L. H. Tjeng, P. J. Ryan and J. W. Freeland, Physical Review Letters **107** (11) (2011).
9. Y. Tokura and N. Nagaosa, Science **288** (5465), 462-468 (2000).
10. M. Imada, A. Fujimori and Y. Tokura, Reviews of Modern Physics **70** (4), 1039-1263 (1998).
11. M. Fiebig, K. Miyano, Y. Tomioka and Y. Tokura, Science **280** (5371), 1925-1928 (1998).
12. E. Dagotto, Science **309** (5732), 257-262 (2005).
13. E. Dagotto, T. Hotta and A. Moreo, Phys. Rep.-Rev. Sec. Phys. Lett. **344** (1-3), 1-153 (2001).
14. J. Tao, D. Niebieskikwiat, M. Varela, W. Luo, M. A. Schofield, Y. Zhu, M. B. Salamon, J. M. Zuo, S. T. Pantelides and S. J. Pennycook, Physical Review Letters **103** (9) (2009).
15. X. Q. Pan, W. D. Kaplan, M. Ruhle and R. E. Newnham, J. Am. Ceram. Soc. **81** (3), 597-605 (1998).
16. Z. Xu, M. C. Kim, J. F. Li and D. Viehland, Philosophical Magazine a-Physics of Condensed Matter Structure Defects and Mechanical Properties **74** (2), 395-406 (1996).
17. X. H. Dai, Z. Xu and D. Viehland, Journal of Applied Physics **79** (2), 1021-1026 (1996).
18. Z. Xu, X. H. Dai, J. F. Li and D. Viehland, Applied Physics Letters **68** (12), 1628-1630 (1996).
19. X. H. Dai, Z. Xu and D. Viehland, J. Am. Ceram. Soc. **79** (7), 1957-1960 (1996).
20. A. E. Glazounov, A. K. Tagantsev and A. J. Bell, Phys. Rev. B **53** (17), 11281-11284 (1996).
21. Y. V. Pershin and M. Di Ventra, Advances in Physics **60** (2), 145-227 (2011).
22. E. Salje and H. L. Zhang, Phase Transitions **82** (6), 452-469 (2009).
23. T. Zhang, Z. T. Song, M. Sun, B. Liu, S. L. Feng and B. Chen, Applied Physics a-Materials Science & Processing **90** (3), 451-455 (2008).
24. J. F. Mitchell, D. N. Argyriou, C. D. Potter, D. G. Hinks, J. D. Jorgensen and S. D. Bader, Phys. Rev. B **54** (9), 6172-6183 (1996).
25. A. Urushibara, Y. Moritomo, T. Arima, A. Asamitsu, G. Kido and Y. Tokura, Physical Review B **51** (20), 14103-14109 (1995).
26. Y. Tokura, M. Kawasaki and N. Nagaosa, Nature Physics **13** (11), 1056-1068 (2017).
27. J. P. Velev, S. S. Jaswal and E. Y. Tsymbal, Philosophical Transactions of the Royal Society a-Mathematical Physical and Engineering Sciences **369** (1948), 3069-3097 (2011).
28. R. J. Zeches, M. D. Rossell, J. X. Zhang, A. J. Hatt, Q. He, C. H. Yang, A. Kumar, C. H. Wang, A. Melville, C. Adamo, G. Sheng, Y. H. Chu, J. F. Ihlefeld, R. Erni, C. Ederer, V. Gopalan, L. Q. Chen, D. G. Schlom, N. A. Spaldin, L. W. Martin and R. Ramesh, Science **326** (5955), 977-980 (2009).

29. A. Haghiri-Gosnet and J. J. J. o. P. D. A. P. Renard, **36** (8), R127 (2003).
30. E. V. Balashova and A. K. Tagantsev, Phys. Rev. B **48** (14), 9979-9986 (1993).
31. W. W. Cao and G. R. Barsch, Phys. Rev. B **41** (7), 4334-4348 (1990).
32. J. Katamura and T. Sakuma, J. Am. Ceram. Soc. **80** (10), 2685-2688 (1997).
33. D. N. Basov, R. D. Averitt and D. Hsieh, Nature Materials **16** (11), 1077-1088 (2017).
34. G. Tenorio, L. Bucio and R. Escudero, Journal of Superconductivity and Novel Magnetism **30** (9), 2381-2386 (2017).
35. Y. Watanabe, Phys. Rev. B **57** (2), 789-804 (1998).
36. L. A. Turkevich and R. A. Klemm, Phys. Rev. B **19** (5), 2520-2539 (1979).
37. J. Hlinka and P. Marton, Phys. Rev. B **74** (10) (2006).
38. B. E. Vugmeister and H. Rabitz, Physical Review B **61** (21), 14448-14453 (2000).
39. M. D. Glinchuk and V. A. Stephanovich, Journal of Applied Physics **85** (3), 1722-1726 (1999).
40. M. D. Glinchuk and V. A. Stephanovich, J. Phys.-Condes. Matter **10** (48), 11081-11094 (1998).
41. K. Binder and J. D. Reger, Adv. Phys. **41** (6), 547-627 (1992).
42. J. Jung, M. A. K. Mohamed, I. Isaac and L. Friedrich, Phys. Rev. B **49** (17), 12188-12199 (1994).
43. M. S. Senn, D. A. Keen, T. C. A. Lucas, J. A. Hriljac and A. L. Goodwin, Physical Review Letters **116** (20) (2016).
44. D. A. Keen and A. L. Goodwin, Nature **521** (7552), 303-309 (2015).
45. A. K. Cheetham, T. D. Bennett, F. X. Coudert and A. L. Goodwin, Dalton Transactions **45** (10), 4113-4126 (2016).
46. I. Levin, V. Krayzman, J. C. Woicik, J. Karapetrova, T. Proffen, M. G. Tucker and I. M. Reaney, Phys. Rev. B **79** (10) (2009).
47. R. J. Zeches, M. D. Rossell, J. X. Zhang, A. J. Hatt, Q. He, C. H. Yang, A. Kumar, C. H. Wang, A. Melville, C. Adamo, G. Sheng, Y. H. Chu, J. F. Ihlefeld, R. Erni, C. Ederer, V. Gopalan, L. Q. Chen, D. G. Schlom, N. A. Spaldin, L. W. Martin and R. Ramesh, Science **326** (5955), 977 (2009).
48. Y. Yang, C.-C. Chen, M. C. Scott, C. Ophus, R. Xu, A. Pryor, L. Wu, F. Sun, W. Theis, J. Zhou, M. Eisenbach, P. R. C. Kent, R. F. Sabirianov, H. Zeng, P. Ercius and J. Miao, Nature **542** (7639), 75-79 (2017).
49. G. Catalan, A. Lubk, A. H. G. Vlooswijk, E. Snoeck, C. Magen, A. Janssens, G. Rispens, G. Rijnders, D. H. A. Blank and B. Noheda, Nature Materials **10** (12), 963-967 (2011).
50. Y. Zhu, C. Ophus, J. Ciston and H. Wang, Acta Materialia **61** (15), 5646-5663 (2013).
51. M. J. Hÿtch, E. Snoeck and R. Kilaas, Ultramicroscopy **74** (3), 131-146 (1998).
52. Y.-M. Kim, J. He, M. D. Biegalski, H. Ambaye, V. Lauter, H. M. Christen, S. T. Pantelides, S. J. Pennycook, S. V. Kalinin and A. Y. Borisevich, Nature Materials **11** (10), 888-894 (2012).
53. D. Cooper, T. Denneulin, N. Bernier, A. Béché and J.-L. Rouvière, Micron **80**, 145-165 (2016).
54. A. Y. Borisevich, H. J. Chang, M. Huijben, M. P. Oxley, S. Okamoto, M. K. Niranjan, J. D. Burton, E. Y. Tsymbal, Y. H. Chu, P. Yu, R. Ramesh, S. V. Kalinin and S. J. Pennycook, Physical Review Letters **105** (8), 087204 (2010).
55. C.-L. Jia, N. Valanoor, J.-Q. He, L. Houben, T. Zhao, R. Ramesh, K. Urban and R. Waser, Nature Materials **6**, 64-69 (2006).
56. C. T. Nelson, B. Winchester, Y. Zhang, S.-J. Kim, A. Melville, C. Adamo, C. M. Folkman, S.-H. Baek, C.-B. Eom, D. G. Schlom, L.-Q. Chen and X. Pan, Nano Letters **11** (2), 828-834 (2011).
57. C.-L. Jia, K. W. Urban, M. Alexe, D. Hesse and I. Vrejoiu, Science **331** (6023), 1420 (2011).
58. A. Y. Borisevich, A. N. Morozovska, Y. M. Kim, D. Leonard, M. P. Oxley, M. D. Biegalski, E. A. Eliseev and S. V. Kalinin, Physical Review Letters **109** (6) (2012).
59. Q. Li, C. T. Nelson, S. L. Hsu, A. R. Damodaran, L. L. Li, A. K. Yadav, M. McCarter, L. W. Martin, R. Ramesh and S. V. Kalinin, Nature Communications **8** (2017).
60. L. Vlcek, M. Ziatdinov, A. Maksov, A. Tselev, A. P. Baddorf, S. V. Kalinin and R. K. Vasudevan, ACS Nano **13** (1), 718-727 (2019).

61. L. Vlcek, A. Maksov, M. H. Pan, R. K. Vasudevan and S. V. Kahnin, Acs Nano **11** (10), 10313-10320 (2017).
62. K. Sohlberg, S. Rashkeev, A. Y. Borisevich, S. J. Pennycook and S. T. Pantelides, Chemphyschem **5** (12), 1893-1897 (2004).
63. Y. M. Kim, A. Morozovska, E. Eliseev, M. P. Oxley, R. Mishra, S. M. Selbach, T. Grande, S. T. Pantelides, S. V. Kalinin and A. Y. Borisevich, Nature Materials **13** (11), 1019-1025 (2014).
64. H. Rønnow, C. Renner, G. Aeppli, T. Kimura and Y. J. N. Tokura, **440** (7087), 1025-1028 (2006).
65. C. W. Bark, P. Sharma, Y. Wang, S. H. Baek, S. Lee, S. Ryu, C. M. Folkman, T. R. Paudel, A. Kumar, S. V. Kalinin, A. Sokolov, E. Y. Tsymbal, M. S. Rzchowski, A. Gruverman and C. B. Eom, Nano Letters **12** (4), 1765-1771 (2012).
66. H. Lu, C. W. Bark, D. E. de los Ojos, J. Alcala, C. B. Eom, G. Catalan and A. Gruverman, Science **336** (6077), 59-61 (2012).
67. R. K. Vasudevan, N. Balke, P. Maksymovych, S. Jesse and S. V. Kalinin, Applied Physics Reviews **4** (2) (2017).
68. A. N. Morozovska, E. A. Eliseev, N. V. Morozovsky and S. V. Kalinin, Phys. Rev. B **95** (19) (2017).
69. A. N. Morozovska, E. A. Eliseev, A. I. Kurchak, N. V. Morozovsky, R. K. Vasudevan, M. V. Strikha and S. V. Kalinin, Phys. Rev. B **96** (24) (2017).
70. S. M. Yang, A. N. Morozovska, R. Kumar, E. A. Eliseev, Y. Cao, L. Mazet, N. Balke, S. Jesse, R. K. Vasudevan, C. Dubourdieu and S. V. Kalinin, Nature Physics **13** (8), 812-818 (2017).
71. L. Itti and P. Baldi, Vision Res **49** (10), 1295-1306 (2009).
72. D. Kan, L. Pálová, V. Anbusathaiah, C. J. Cheng, S. Fujino, V. Nagarajan, K. M. Rabe and I. Takeuchi, Advanced Functional Materials **20** (7), 1108-1115 (2010).
73. S. Karimi, I. M. Reaney, Y. Han, J. Pokorny and I. Sterianou, Journal of Materials Science **44** (19), 5102-5112 (2009).
74. R. Guo, K. M. Nair, W. Wong-Ng, A. Bhalla, D. Viehland, D. Suvorov, C. Wu and S.-I. Hirano, *Morphotropic Phase Boundary Perovskites, High Strain Piezoelectrics, and Dielectric Ceramics,*. (American Ceramic Society, 2003).
75. A. G. Khachaturyan, Philosophical Magazine **90** (1-4), 37-60 (2010).
76. C. J. Cheng, D. Kan, S. H. Lim, W. R. McKenzie, P. R. Munroe, L. G. Salamanca-Riba, R. L. Withers, I. Takeuchi and V. Nagarajan, Physical Review B **80** (1), 014109 (2009).
77. A. Y. Borisevich, E. A. Eliseev, A. N. Morozovska, C. J. Cheng, J. Y. Lin, Y. H. Chu, D. Kan, I. Takeuchi, V. Nagarajan and S. V. Kalinin, Nature Communications **3** (1), 775 (2012).
78. A. Gruverman, O. Auciello, R. Ramesh and H. Tokumoto, Nanotechnology **8**, A38-A43 (1997).
79. N. Balke, I. Bdikin, S. V. Kalinin and A. L. Kholkin, J. Am. Ceram. Soc. **92** (8), 1629-1647 (2009).
80. M. Ziatdinov, C. Nelson, X. Zhang, R. Vasudevan, E. Eliseev, A. N. Morozovska, I. Takeuchi and S. V. Kalinin, arXiv preprint arXiv:2002.04245 (2020).
81. C. Ophus, C. T. Nelson and J. Ciston, Ultramicroscopy **162**, 1-9 (2016).
82. M. Ziatdinov, C. T. Nelson, X. Zhang, R. K. Vasudevan, E. Eliseev, A. N. Morozovska, I. Takeuchi and S. V. Kalinin, npj Computational Materials **6** (1), 127 (2020).
83. C. E. Rasmussen and C. K. I. Williams, *Gaussian Processes for Machine Learning (Adaptive Computation and Machine Learning)*. (The MIT Press, 2005).
84. O. Martin, *Bayesian Analysis with Python: Introduction to statistical modeling and probabilistic programming using PyMC3 and ArviZ, 2nd Edition*. (Packt Publishing, 2018).
85. B. Lambert, *A Student's Guide to Bayesian Statistics*. (SAGE Publications Ltd; 1 edition, 2018).
86. J. Kruschke, *Doing Bayesian Data Analysis: A Tutorial with R, JAGS, and Stan*. (Academic Press; 2 edition, 2014).
87. J. Quiñonero-Candela and C. E. Rasmussen, J. Mach. Learn. Res. **6** (Dec), 1939-1959 (2005).

88. E. Bingham, J. P. Chen, M. Jankowiak, F. Obermeyer, N. Pradhan, T. Karaletsos, R. Singh, P. Szerlip, P. Horsfall and N. D. Goodman, The Journal of Machine Learning Research **20** (1), 973-978 (2019).
89. M. Skurichina and R. P. W. Duin, Pattern Recognit. **31** (7), 909-930 (1998).
90. D. R. Hardoon, S. Szedmak and J. Shawe-Taylor, Neural Computation **16** (12), 2639-2664 (2004).
91. K. Pearson, Proceedings of the Royal Society of London **58**, 240-242 (1895).